# Realism vs. Constructivism in Contemporary Physics: The Impact of the Debate on the Understanding of Quantum Theory and its Instructional Process

VASSILIOS  KARAKOSTAS  and  PANDORA  HADZIDAKI

*Department of Philosophy and History of Science, University of Athens,
Athens 157 71, Greece*; E-mail: karakost@phs.uoa.gr, hadzidaki@otenet.gr

**Abstract:** In the present study we attempt to incorporate the philosophical dialogue about physical reality into the instructional process of quantum mechanics. Taking into account that both scientific realism and constructivism represent, on the basis of a rather broad spectrum, prevalent philosophical currents in the domain of science education, the compatibility of their essential commitments is examined against the conceptual structure of quantum theory. It is argued in this respect that the objects of science do not simply constitute 'personal constructions' of the human mind for interpreting nature, as individualist constructivist consider, neither do they form products of a 'social construction', as sociological constructivist assume; on the contrary, they reflect objective structural aspects of the physical world. A realist interpretation of quantum mechanics, we suggest, is not only possible but also necessary for revealing the inner meaning of the theory's scientific content. It is pointed out, however, that a viable realist interpretation of quantum theory requires the abandonment or radical revision of the classical conception of physical reality and its traditional metaphysical presuppositions. To this end, we put forward an alternative to traditional realism interpretative scheme, that is in harmony with the findings of present-day quantum theory, and which, if adequately introduced into the instructional process of contemporary physics, is expected to promote the conceptual reconstruction of learners towards an appropriate view of nature.

## 1.  Introduction

The development of quantum mechanics (QM) has not only suggested a radically new scientific viewpoint for the physical world, but has also formed the conceptual basis for the proper interpretation of a wide range of physical phenomena. This fact forces school curricula to introduce QM topics at an early stage of the instructional process. However, empirical studies concerning science education reveal the following important aspects in relation to QM teaching: Firstly, there is strong evidence that traditional teaching methods introduce certain aspects of subatomic phenomena in ways that usually cause an *awkward intermixture* of the conceptual systems of classical physics (CP) and QM. Thus, students' cognitive structures are generally characterized by a rather *classical perception* of quantum physics, which incorporates, in an incompatible manner, "elements of both mechanistic and quantum ideas" (Ireson 1999, p. 78). Such elements transfer, for example, macroscopic attributes to the submicroscopic objects or diffuse uncritically the deterministic mode of classical thinking and reasoning into QM context (Kalkanis, Hadzidaki & Stavrou 2003, pp. 267-268). Secondly, the so formed cognitive structures of students seem to exhibit a strong *competitive action* with respect to a more literal QM learning. Specifically, students tend to intuitively assimilate the newly considered quantum mechanical concepts into categories and modes of thinking that are deeply rooted into the conceptual network of CP. Thus, learners' misconceptions, appearing as "the contrasting point of reference for each new idea" (Petri & Niedderer 1998, p. 1079), form a species of students' "stable knowledge" (Fischler & Peuchert 1999, p. 397), which seems to get stabilized in an inappropriate manner even after a specialized educational training.

These findings raise, in our opinion, the demand of framing educational strategies that may allow learners to gain a sufficiently informed and, above all, scientifically pertinent insight to QM worldview. In order to respond to this requirement, we designed a project that aims at enriching the *epistemological* and *cognitive* foundation of a qualitative instructional approach to contemporary physics. Within this context of inquiry, the cognitive component of the present study is shaped by certain conclusive results of cognitive psychology that emphasize the influence of learners' *initial ontological categorizations* on knowledge acquisition. It has been claimed, for instance, that "the mismatch or incompatibility between the categorical representations that students bring to an instructional context and the ontological category to which the science concept truly belongs" (Chi, Slotta & Leeuw 1994, p. 34) contributes, in an essential way, to the formation of learners' misconceptions. We indeed accept, in this respect, that a conceptual reconstruction of learners, with regard to physics instruction, presupposes an explicit discrimination of the ontological/conceptual assumptions underlying the theoretical structures of classical and quantum physics. Thus, we attempt in this study to incorporate the philosophical dialogue about 'physical reality' into QM instruction, as follows: Firstly, we determine the key-differentiations between the conceptual structures of classical and quantum mechanics; as key-differentiations, we mean those that deliberately impose a different way of viewing physical phenomena. Secondly, taking into account that both *scientific realism* and *constructivism* represent, on the basis of a rather broad spectrum, prevalent philosophical currents in the domain of science education, we examine the *compatibility* of their essential commitments (of an epistemological and ontological nature) with QM scientific content. By means of such a critical analysis, we finally put forward a suitable interpretative scheme that is fully harmonized with present-day quantum theory. This scheme, adequately introduced into QM instruction, is expected to effectively support students' conceptual change.

## 2. Classical World vs. Quantum World

Classical physics (CP) is essentially *atomistic* in character. It portrays a view of the world in terms of analyzable, separately existing but interacting self-contained parts. CP is also *reductionistic*. It aims in explaining the whole of forms of physical existence, of structures and relations of the natural world in terms of a salient set of elementary material objects linked by forces. CP (and practically any experimental science) is further based on the *Cartesian dualism* of 'res cogitans' ('thinking substance') and 'res extensa' ('extended substance'), proclaiming a radical separation of an objective external world from the knowing subject that allows no possible intermediary.

In fact, the whole edifice of CP — be it point-like analytic, statistical, or field theoretic — obeys a *separability principle* that can be expressed schematically as follows:

*Separability Principle*: "The states of any spatio-temporally separated subsystems $S_1$, $S_2$, ..., $S_N$ of a compound system S are *individually well defined* and the states of the compound system are *wholly* and *completely determined* by them and their physical interactions including their spatio-temporal relations" (Karakostas 2004a, p. 284 and references therein).

The aforementioned separability principle delimits actually the fact, upon which the whole CP is founded, that any compound physical system of a classical universe can be conceived of as consisting of separable, distinct parts interacting by means of forces, which are encoded in the Hamiltonian function of the overall system, and that, if the full Hamiltonian is known, *maximal knowledge of the values of the physical quantities pertaining to each one of these parts yields an exhaustive knowledge of the whole compound system*.

The notion of separability has been viewed within the framework of CP as a principal condition of our conception of the world, a condition that characterizes all our thinking in acknowledging the physical identity of distant things, the "mutually independent existence (the 'being thus')" of spatio-temporally separated systems (Einstein 1948/1971, p. 169). The primary implicit assumption pertaining to this view is a presumed *absolute kinematic independence* between the knowing subject (the physical scientist) and the object of knowledge, or equivalently, between the measuring system (as an extension of the knowing subject) and the system under measurement. The idealization of the kinematically independent behaviour of a physical system is possible in CP both due to the Cartesian-product structure of phase space, namely, the state space of classical theories, and the absence of genuine indeterminism in the course of events or of an element of chance in the measurement process. During the act of measurement a classical system conserves its identity. Successive measurements of physical quantities, like position and momentum that define the state of a classical system, can be performed to any degree of accuracy and the results combined can completely determine the state of the system before and after the measurement interaction, since its effect, if not eliminable, takes place continuously in the system's state space and is therefore predictable in principle. Consequently, classical physical quantities or properties are taken to obey a so-called '*possessed values*' principle, in the sense that, the values of classical properties are considered as being possessed by the object itself independently of any measurement act. That is, the properties possessed by an object depend in no way on the relations obtaining between it and a possible experimental context used to bring these properties about. No qualitatively new elements of reality are produced by the interaction of a classical system with the measuring apparatus. The act of measurement in CP is passive; it simply reveals a fact which has already occurred. In other words, a substantial distinction between *potential* and *actual* existence is rendered obsolete in classical mechanics. Within the domain of the latter, all that is potentially possible is also actually realised in the course of time, independently of any measuring interventions. It should be noted, in this respect, that this is hardly the case in the quantum theory of the measurement process (Karakostas 1994, Karakostas & Dickson 1995).

In contrast to CP, standard QM systematically violates the conception of separability. From a formal point of view, the source of its defiance is due to the tensor-product structure of Hilbert-space QM and the superposition principle of states, which incorporates a kind of *objective indefiniteness* for the numerical values of *any* observable belonging to a superposed state. The generic phenomenon of quantum nonseparability, experimentally confirmed for the first time in the early 1980s, precludes in a novel way the possibility of defining individual objects independently of the conditions under which their behaviour is manifested. Even in the simplest possible case of a compound system S consisting of just two subsystems $S_1$ and $S_2$ that have interacted at some time in the past, the compound system should be treated as a *nonseparable, entangled* system, however large is the distance among $S_1$ and $S_2$. In such a case, it is not permissible to consider them individually as distinct entities enjoying intertemporal identity. The global character of their behaviour precludes any description or any explanation in terms of individual systems, each with its own well-defined state or predetermined physical properties. Only the compound system S, as a whole, is assigned a well-defined (nonseparable) pure state. Therefore, when a compound system such as S is in an entangled state, namely a superposition of pure states of tensor-product forms, *maximal knowledge of the whole system does not allow maximal knowledge of its component parts*, a circumstance with no precedence in CP. In a paper related to the Einstein-Podolsky-Rosen argument, Schrödinger explicitly anticipated this counterintuitive state of affairs:

''When two systems, of which we know the states by their respective representations, enter into temporary physical interaction due to known forces between them, and then after a time of mutual influence the systems separate again, then *they can no longer be described in the same way as before, viz. by endowing each of them with a representative of its own. ...* I would not call that one but rather *the characteristic trait of quantum mechanics, the one that enforces its entire departure from classical lines of thought*'' (Schrödinger 1935/1983, p. 161).

The phenomenon of quantum nonseparability undeniably reveals the holistic character of entangled quantum systems. Quantum mechanics is the first — and up to day the only — logically consistent, mathematically formulated and empirically well-confirmed theory, which incorporates as its basic feature that the 'whole' is, in a non-trivial way, more than the sum of its 'parts' including their spatiotemporal relations and physical interactions. Contrary to the situation in CP, when considering an entangled compound system, 'whole' and 'parts' are dynamically related in such a way that their *bi-directional reduction* is, in principle, impossible (e.g., Karakostas 2004a).

From a foundational viewpoint of quantum theory, quantum mechanical nonseparability and the sense of quantum holism arising out of it refer to a context-independent, or in d' Espagnat's (1995) scheme, observer- or mind-independent reality. The latter is operationally inaccessible. It pertains to the domain of entangled correlations, potentialities and quantum superpositions obeying a non-Boolean logical structure. Here the notion of an object, whose aspects may result in intersubjective agreement, enjoys no a priori meaning independently of the phenomenon into which is embedded. In QM in order to be able to speak meaningfully about an object, to obtain any kind of description, or refer to experimentally accessible facts the underlying wholeness of nature should be decomposed into interacting but disentangled subsystems, namely into 'measured objects' and 'detached observers' (measuring apparata) with no (or insignificantly so) entangled correlations among them. This subject-object separation is sometimes metaphorically called the Heisenberg cut (e.g., Heisenberg 1958, p. 116).

The presuppositions of the latter are automatically satisfied in CP, in consonance with the aforementioned separability principle. In a nonseparable theory like QM, however, the concept of the Heisenberg cut acquires the status of a methodological regulative principle through which access to empirical reality is rendered possible. The innovation of the Heisenberg cut, and the associated separation of a quantum object from its environment, is mandatory for the description of measurements. It is, in fact, necessary for the operational account of any directly observable pattern of empirical reality. The very possibility of devising and repeating a controllable experimental procedure presupposes the existence of such a subject-object separation. Without it the concrete world of material facts and data would be ineligible; it would be conceived in a totally entangled manner. In this sense, a physical system may account as an experimental or a measuring device only if it is not holistically correlated or entangled with the object under measurement.

Consequently, any atomic fact or event that 'happens' is raised at the observational level only in conjunction with the specification of an experimental arrangement — an experimental context that conforms to a Boolean domain of discourse — namely to a set of observables co-measurable by that context. In other words, there cannot be well-defined events in QM unless a specific set of co-measurable observables has been singled out for the system-experimental context whole. For, in the domain of QM, one cannot assume, with no falling into contradictions, that observed objects enjoy a separate well-defined identity irrespective of any particular context. One cannot assign, in a consistent manner, definite sharp values to all quantum mechanical observables pertaining to a microscopic object, in particular to pairs of incompatible observables, independently of the measurement context actually specified. In terms of the structural component of QM, this is due to functional relationship constraints that govern the algebra of quantum mechanical observables, as revealed by the Kochen-Specker (1967) theorem and its recent investigations. In fact, any attempt of simultaneously attributing sharp values to all observables of a microscopic object forces the quantum statistical distribution of value assignment into the pattern of a classical distribution, thus leading, for instance, to contradictions of the GHZ-type (e.g., Greenberger, Horne, Shimony & Zeilinger 1990; for a more recent discussion see Mermin 1998; see also Forge 2003).

The context dependence of the attribution of values to quantum observables has been expressly mentioned by Bell in his pioneering analysis on the interpretative status of QM:

"It was tacitly assumed that measurement of an observable must yield the same value independently of what other measurements may be made ... There is no a priori reason to believe that the result should be the same. The result of an observation may reasonably depend not only on the state of the system ... but also on the complete disposition of the apparatus" (Bell 1966, p. 451).

The context dependence of measurement results appears even earlier in the writings of Bohr who repeatedly emphasized

"the impossibility of any sharp distinction between the beahaviour of atomic objects and the interaction with the measuring instruments which serve to define the conditions under which the phenomena appear" (Bohr 1949, p. 210).

This feature of contextuality concerning the attribution of properties in QM is also present in Bohm's ontological interpretation of the theory by clearly putting forward that

"quantum properties cannot be said to belong to the observed system alone and, more generally, that such properties have no meaning apart from the total context which is relevant in any particular situation. In this case, this includes the overall experimental arrangement so that we can say that measurement is context dependent" (Bohm and Hiley 1993, p. 108).

This state of affairs reflects most clearly the unreliability of the common-sensical 'possessed values' principle of CP, according to which, values of physical quantities are regarded as being possessed by an object independently of any measurement context. The classical realist underpinning of such an assumption has been shown to be *incompatible* with the structure of the algebra of quantum mechanical observables. Well defined values of quantum observables can be regarded as pertaining to an object of our interest only within a framework involving the experimental conditions. The latter provide the necessary conditions whereby we make meaningful statements that the properties we attribute to quantum objects are part of physical reality. Consequent upon that the nature of quantum objects is a *context-dependent* issue with the experimental procedure supplying the physical context, the necessary frame for their being. Within the nonseparable structure of QM, the role of objects is not considered as being 'behind' the phenomena, but 'in' the phenomena (e.g., v. Weizcäcker 1971, p. 28). This seems to be reminiscent of Kant's view that the concept of an object is a condition of the possibility of its being experienced. However this may be, the meaning of the term reality in the quantum realm cannot be considered to be determined by what physical objects really are in themeselves. Quantum mechanics describes material reality in a substantially context-dependent manner. The classical idealization of sharply individuated objects possessing intrinsic properties and having an independent reality of their own breaks down in the quantum domain.

## 3. Classical Scientific Realism Embodies the Ontology of a Classical World

The basic presupposition that shapes classical scientific realism (CSR) lies along the thesis of 'metaphysical independence': the putative objects of scientific description and explanation belong to a world *independent* of the scientist, independent not in the trivial sense of being there whether or not human observers exist, but in the sense of being *the way it is* whether or not it is observed and regardless of the acts or operations of the observer. Thus, in CSR generally appear two distinct, albeit interrelated, dimensions: the 'existence dimension', which constitutes an indispensable premise of *every* realistic interpretation of the physical world, and the 'independence dimension', which constitutes an additional essential ingredient of CSR's ontological scheme (cf. Devitt 1997, pp. 14-22). Clearly, the 'independence dimension', in order to be satisfied, demands, in addition to the 'existence dimension', a clear distinction between the perceiving subject and the observed object. As Feyerabend appositely remarks, a realist who adopts CSR "will insist on the separation between subject and object and he will try to restore it wherever research seems to have found fault with it" (Feyerabend 1981, p. 72).

As already noted in section 2, the logical structure of CP, including as an inherent element the principle of separability, may naturally seem to secure the subject – object total separation. It is not surprising then that CP is receptive to an interpretational account nicely harmonized with CSR's ontological commitments. Nonetheless, this fact may be reliably established, if we compare the underlying assumptions of CP with the following basic theses of CSR — namely, the 'metaphysical', the 'semantic' and the 'epistemic' — on which all variants of traditional realism would agree to a reasonable extent.

### 3.1. THE 'METAPHYSICAL' THESIS OF CSR

The separability principle permits one to legitimately consider that the 'cut' of a physical whole, not being affected by the intervention of the cognizing subject, leads to well-defined individual parts, which retain their own particular identity *intact*. This feature of CP's formal structure implicitly implies the notion of *ontological reductionism,* according to which, a salient subset of the natural kind of entities inhabiting the world, together with their properties and spatio-temporal relations they enter into, *fix* or *determine*, through a series of successive reductions, the nature and behaviour of the universe as a whole. Thus, within the framework of CP, the upholders of CSR are entitled to consider that "the world exists and is organized *independently* of us, our language, and our methods of inquiry" (Hwang 1996, p. 345). This statement presents the 'metaphysical' thesis of CSR.

## 3.2. THE 'SEMANTIC' THESIS OF CSR

The separability principle also promotes the idea of *epistemological reductionism*: as the 'cut' of a whole causes no loss of knowledge, science can attain *a complete* understanding of the whole in terms of its parts. This, by extension, amounts to the fact that once maximal knowledge of the individual physical entities constituting the world is attained, the physical world itself, as a unified whole, is ideally completely knowable; it forms an *already* structured whole that remains to be discovered by scientific inquiry. But, is indeed science capable of describing physical entities as they 'really are'? The answer within the framework of CP could be affirmative, without being open to disproval, since the possibility of detaching, during the measurement process, a physical entity from its environment permits the following inferences: *Firstly*, the measured values can be viewed as corresponding to physical quantities *possessed* by the entity itself; these quantities can therefore be viewed as *intrinsic properties* of the entity under consideration. *Secondly*, the attributed properties, remaining qualitatively *unchangeable* during the measuring process, can be regarded as constituent elements of the entity's *stable* identity. And, *thirdly*, the process of measurement or observation, causing no loss of knowledge, can be considered as offering the necessary data for the accurate representation of the observed entity, as it 'really is'. Now, since every physical entity (observable or unobservable) may be thought of to be identifiable in this way, CP permits substantial ontological commitments to the unobservable entities posited by scientific theories. Thus, it further allows the adherents of CSR to rigorously assert that "theoretical terms in scientific theories, i.e., nonobservational terms, should be thought of as putatively *referring* expressions" (Boyd 1983, p. 195). This statement presents the 'semantic' thesis of CSR, according to which, scientific terms are assigned a well-defined meaning by virtue of their *genuine reference* to entities that populate the world.

On this issue, we have to pay also attention on the fact that classical statistical considerations in no way shake the semantic thesis of CSR: not questioning the individual existence of physical entities, they simply reflect, due to the complexity of the situation, our epistemic ignorance or lack of knowledge, probably temporary, about the finer features of a grossly complex phenomenon. Thus, it is often thought that, within the classical universe, the ultimate attributes of physical objects 'as they really are' can and will (perhaps, asymptotically in time only) be known. This is an assumption that perfectly fits to the claim of CSR that "the historical progress of mature sciences is largely a matter of successively more accurate approximations to the truth about both observables and unobservables" (Boyd 1983, p. 195).

### 3.3. THE 'EPISTEMIC' THESIS OF CSR

The validity of the separability principle in CP gradually portrayed a common sensical realistic view of the external world as being completely detached from the perceiving subject. As we saw, given this pattern, the rationale behind CP implied that the observed behaviour of physical entities in the world corresponds to the ontology presumed by the theory. This framework provided further the ground for CSR to adopt as a notion of scientific truth a *representational correspondence* between the way the world actually is and the way we observe the world to be. In so far as classical concepts can be consistently viewed as referring to entities of the 'real world as it truly is', classical physical science, as a whole, may be, and indeed is systematically viewed by adherents of CSR as providing an approximately accurate representation of the world as 'it really is'. In this sense, the configuration and development of CP has been frequently used as a means for supporting the claim that "the laws and theories embodied in our actual theoretical tradition are approximately true of the world" (Boyd 1983, p. 207). This statement presents the 'epistemic' thesis of CSR.

It should be noted in this respect that the latter thesis of traditional realism guarantees, in what may be called in d' Espagnat's (1995, p. 22) terms, a 'strong' version of objectivity in science. The exact meaning of the qualificative 'strong' is appropriately expressed by Dumett, when he portrays scientific realism as the belief, according to which, "there are statements that possess an objective truth-value, independently of our means of knowing it: they are true or false in virtue of a reality existing *independently* of us" (Dumett 1978, p. 210; see also Psillos 2000, p. 713). Although, within the framework of CP, the general validity of Dumett's assertion can neither be actually proved or disproved, nonetheless, classical physics may be classified as a strongly objective theory, since all its basic laws are couched in terms of strongly objective propositions; strongly objective, in the sense that, being formulated under the presupposition of subject-object separation, they respond to Dumett's aforementioned prerequisite of 'not being dependent on our means of knowing'. That is, propositions in CP are considered as being either true or false in virtue of a reality existing independently of us, and, thus, as possessing an objective truth value, regardless of our means of exploring and certifying it.

The preceding analysis illustrates the fact that CSR's philosophical package is suitably equipped to offer a putative ontological interpretation of classical physical science. However, if one considers that a 'quantum revolution' actually happened, he ought to admit that its very essence lies on the fact that QM incorporates a form of holism absent from classical physics (see section 2). This fact, of course, questions in a direct way the very initial premise of CSR, namely, the subject–object separability. Hence, the ontological scheme of CSR becomes, as will be shown in section 5, at least unconvincing. But accepting CSR's weakness in offering a realistic view of the physical world that is compatible with QM, it does not entail that there are no arguments whatsoever for any form of realism in the quantum realm. In what follows we attempt indeed to show that, if one wants to adequately describe the quantum image of the world, a realistic interpretation of QM is not only possible, but also necessary. Thus, for instance, contrary to the multifarious constructivist views on science, we provide compelling reasons, for answering *yes* to the question 'do we need an external reality?'

## 4. An Abridged Guide to Constructivist Perspectives on Science

Constructivism, as portrayed by its adherents, "is the idea that we construct our own world rather than it being determined by an outside reality" (Riegler 2001, p.1). Indeed, a common ground among constructivists of different persuasion lies in a commitment to the idea that knowledge *is actively built up* by the cognizing subject. But, whereas *individualistic* constructivism (which is most clearly enunciated by *radical* constructivism) focuses on *the biological/psychological* mechanisms that lead to knowledge construction, *sociological* constructivism focuses on *the social factors* that influence learning. It is evident that a thorough appraisal of the diverse constructivist variants surpasses the object of this study. Additionally, we take for granted the fact that knowledge acquisition is never an independent function, but always relative to its historical and cultural context. Thus, we limit ourselves to outlining only those constructivist considerations (not necessarily representing all constructivist versions), which in our opinion oppose a compatible with QM understanding of the physical world.

4.1. KNOWLEDGE AND TRUTH

To begin with, let us briefly consider certain fundamental assumptions of individualistic constructivism. The first issue a constructivist theory of cognition ought to elucidate concerns of course the raw materials on which knowledge is constructed. On this issue, von Glaserfeld, an eminent representative of radical constructivism, gives a categorical answer: "from the constructivist point of view, the subject cannot transcend the limits of individual experience" (quoted in Matthews 1998, p. 2). This statement presents the keystone of constructivist epistemology, which conclusively asserts that "*the only* tools available to a 'knower' are the senses … [through which] the individual builds *a picture* of the world" (Lorsbach & Tobin 1992, p. 5). What is more, the so formed mental pictures do not shape an 'external' to the subject world, but the *distinct* personal reality of each individual. And this of course entails, in its turn, that the responsibility for the gained knowledge lies with the constructor; it cannot be shifted to a pre-existing world. As Glanville confesses, "reality is what I sense, as I sense it, when I'm being honest about it" (Glanville 2001, p. 54).

In this way, individualistic constructivism estranges the cognizing subject from the external world. Cognition is not considered as aiming at the discovery and investigation of an 'independent' world; it is viewed as a 'tool' that exclusively serves the adaptation of the subject to the world as it is experienced. From this perspective, 'knowledge' acquires an entirely new meaning. In the expression of von Glaserfeld, the word 'knowledge' refers to conceptual structures that epistemic agents, given the range of *present experience*, within their tradition of thought and language, consider *viable* (von Glaserfeld 1995, p. 14).

[Furthermore] concepts have to be *individually* built up by reflective abstraction; and reflective abstraction is not a matter of looking closer but at operating mentally in a way that happens to be *compatible* with *the perceptual material* at hand (ibid., p. 184).

To say it briefly, 'knowledge' signifies nothing more than an adequate organization of the experiential world, which makes the cognizing subject capable to effectively manipulate its perceptual experience.

It is evident that such insights, precluding any external point of reference, have impacts on knowledge evaluation. Indeed, the ascertainment that "for constructivists there are no structures other than those which the knower forms by its own activity" (von Glaserfeld, quoted in Matthews 1994, p. 141) yields unavoidably the conclusion that "there is no mind-independent yardstick against which to measure the quality of any solution" (De Zeeuw 2001, p. 92). Hence, knowledge claims should not be evaluated by reference to a supposed 'external' world, but only by reference to their internal consistency and personal utility. This is precisely the reason that leads Glaserfeld to suggest the substitution of the notion of "truth" by the notion of "viability" or "functional fit": knowledge claims are appraised as "true", if they "functionally fit" into the subject's experiential world; and to find a "fit" simply means not to notice any discrepancies. This functional adaptation of 'knowledge' to experience is what finally secures the intended "viability" (von Glaserfeld 2001, pp. 39-40).

4.2. OBJECT AND OBJECTIVITY

In accordance with the constructivist view, the notion of 'object', far from indicating any kind of 'existence', it explicitly refers to a strictly *personal* construction of the cognizing subject. Specifically, "any item of the furniture of someone's experiential world can be called an 'object' " (von Glaserfeld 2001, p. 36). From this point of view, the supposition that "the objects one has isolated in his experience are identical with those others have formed … is an illusion" (ibid., p. 37). This of course deprives language of any rigorous criterion of objectivity; its physical-object statements, being dependent upon elements that are derived from personal experience, cannot be considered to reveal attributes of the objects as they factually are. Incorporating concepts whose meaning is highly associated with the individual experience of the cognizing subject, these statements form at the end a personal-specific description of the world. Conclusively, for constructivists the term 'objectivity' "shows no more than a *relative compatibility* of concepts" in situations where individuals have had occasion to compare their "*individual uses* of the particular words" (ibid., p. 37).

4.3. THE STATUS OF SCIENTIFIC KNOWLEDGE — THE REDUNDANCY OF AN INDEPENDENT REALITY

From the viewpoint of radical constructivism, science, being a human enterprise, is amenable, by its very nature, to human limitations. It is then naturally inferred on constructivist grounds that "science cannot transcend [just as individuals cannot] the domain of experience" (von Glaserfeld 2001, p. 31). This statement, indicating that there is no essential differentiation between personal and scientific knowledge, permits, for instance, Staver to assert that "for constructivists, observations, objects, events, data, laws and theory *do not exist* independent of observers. The lawful and certain nature of natural phenomena is a property of *us*, those who describe, not of nature, what is described" (Staver 1998, p. 503). Accordingly, by virtue of the preceding premise, one may argue that "scientific theories *are derived* from human experience and formulated in terms of human concepts" (von Glaserfeld 2001, p. 41).

In the framework now of social constructivism, if one accepts that the term 'knowledge' means no more than "what is collectively endorsed" (Bloor 1991, p. 5), he will probably come to the conclusion that "the natural world has a *small* or *non-existent* role in the construction of scientific knowledge" (Collins 1981, p. 3). Or, in a weaker form, one can postulate that "scientific knowledge is symbolic in nature and socially negotiated. The objects of science *are not* the phenomena of nature but constructs advanced by the scientific community to interpret nature" (Driver et al. 1994, p. 5). It is worth remarking that both views of constructivism eliminate, or at least downplay, the role of the natural world in the construction of scientific knowledge.

It is evident that the foregoing considerations lead most versions of constructivism to ultimately conclude that the very word 'existence' has no meaning in itself. It does acquire meaning only by referring to individuals or human communities. The acknowledgement of this fact renders subsequently the notion of 'external' physical reality useless and therefore redundant. As Riegler puts it, within the constructivist framework, "an external reality is neither rejected nor confirmed, it must be *irrelevant*" (Riegler 2001, p. 5).

**5.  Is Quantum Mechanical Reality Really a Human Construct ?**

As already analyzed in section 3, the ordinary notion of an 'external' reality is in harmony with the underlying assumptions of classical physics. Given, however, the substantial conceptual differentiation between the frameworks of classical and quantum mechanics, could this notion be regarded as irrelevant within the domain of the latter? Although no *logically binding* arguments can be provided in favour of a realistic interpretation of QM, nonetheless, one can list a number of arguments that have quite a strong plausibility value. To this end any appeal to the arguments of classical scientific realism (CSR) should be appropriately circumscribed, since, in view of present-day quantum theory, many of the latter appear, as shown below, at least shaky. Let us then discuss the following points.

5.1. INDEPENDENT REALITY AND CONTEMPORARY PHYSICS

To begin with, we recall that an indispensable prerequisite of the measurement or observational process is the complete discrimination between the observing subject and the object of observation. Thus, fundamental QM, being a nonseparable theory that denies in principle the possibility of such discrimination, can hardly be conceived as a theory exclusively 'derived from human experience', according to the suggestion of Glaserfeld. On the contrary, as Bohr stressed, the consistent formulation and interpretation of QM has been historically attained only *after* "the recognition of physical laws which lie *outside* the domain of human ordinary experience and which present difficulties to human modes of perception" (Bohr 1929, p. 5).

On the other hand, QM undeniably supports the claim that a complete knowledge of the physical word, as 'it really is', is scientifically unfeasible. Indeed, the phenomenon of quantum nonseparability, deeply entrenched into the logical structure of the theory (see section 2), is definitely at variance with the notion of 'ontological reductionism' as implied by CP and advocated by the 'metaphysical' thesis of CSR (see section 3). Quantum nonseparability promotes instead the idea of 'ontological holism' suggesting that the functioning of the physical world *cannot just be reduced* to that of its constituents thought of as a collection of interacting but separately existing localized objects. It should be emphasized, in this connection, that the value of the reductionistic concept as a working hypothesis or as a *methodological* tool of analysis and research is not jeopardized at this point, but, *ontologically*, it can no longer be regarded as a true code of the actual character of the physical world and its contents. For, any compound entangled system has properties which, in a clearly specifiable sense, characterise the whole system but are neither reducible to nor implied by or derived from any combination of local properties of its parts (Karakostas 2003). The establishment of this fact, directly questioning the 'epistemic' thesis of CSR, may seem to support in part the constructivist perspective. However, there is more to say. As the birth process of QM taught us, there is something 'external' to mind structures, which, by *resisting* human attempts to organize and conceptually represent experience, shows off the priority of its 'existence' with respect to human 'knowledge' and 'experience'. In this sense, QM may be thought of as being at least compatible with a "negative" form of realism, stating that "we could call 'real', what, in resisting our constructions, displays its *independent existence* in respect to them" (Baltas 1997, p. 79). We note that this statement expresses a minimal realist thesis, since, it unquestionably adopts the 'existence' dimension of CSR, whilst, it omits its 'independence' dimension (see section 3). Thus, it permits one to recognize that "we are always already 'in' the world and it makes no sense to talk of it from some Archimedean standpoint, as if it where possible to erect ourselves 'outside' " (ibid., p. 76).

This assumption, reminding us the ever-pervasive philosophical difficulty of distinguishing between subject and object, expresses at the level of contemporary physics the very essence of quantum nonseparability. The latter reveals in a conclusive way that — in contrast to an immutable and universal 'view from nowhere' of the classical paradigm — the hunt of a 'bird's-eye' transcendental frame of reference for describing physical reality is in vain (Karakostas 2004a). If then, on the one hand, one accepts the existence of a 'mind-independent' physical world, but, on the other, he recognizes its limited knowability, he may conclude — as d' Espagnat (1995) does — that this world, as regards its holistic nonseparable nature, constitutes a "veiled reality" that is not scientifically completely *accessible*. Of course, a constructivist could raise the objection that this notion of 'reality' appears to be 'useless' and therefore 'redundant'. To answer this, it suffices to show that QM, apart from establishing the weakness of science to fully describe the detailed features of mind-independent reality, it also offers the means for adequately appraising the informational content that this reality *itself* contributes to our knowledge. Let us then proceed along the following line of thought.

## 5.2. THE CONTEXT–DEPENDENCE OF QUANTUM OBJECTS — AN INDEPENDENT REALITY IS STILL UNASSAILABLE

We first recall that the epistemic deconstruction of a compound quantum system, in the face of an intended measurement (the so-called Heisenberg cut, see section 2), provides a level of description to which one can associate a *separable* concept of reality whose elements are commonly experienced as distinct, well-defined objects. The reader should also be reminded, however, that the so-produced objects acquire their own specific 'identity' — represented quantum mechanically by their own separate, contextual state — only within a specific context of investigation which determines the possibilities of their 'being'. Consequently, the said 'objects', being *context-dependent*, cannot be conceived of as 'things in themselves', as 'absolute' building blocks of reality (see also Pauri 2003). Instead, they represent carriers of patterns or properties which arise in interaction with their experimental environment, or more generally, with the rest of the world; the nature of their existence depends on the context into which they are embedded and on the abstractions we are forced to make in any scientific discussion. Hence, instead of picturing entities populating the mind-independent reality, they determine *the possible manifestations* of these entities within a concrete experimental context. In this sense, contextual quantum objects may be viewed as 'constructed', 'phenomenal' entities brought forward by the theory. They do not constitute, however, arbitrary 'personal constructions' of the human mind (as individualist constructivist consider) neither do they form products of a 'social construction' (as sociological constructivist assume). By reflecting the *inherent potentialities* of a quantum entity with respect to a certain pre-selected experimental arrangement, the resulting contextual object may be thought of as a 'construction', as an abstraction-dependent existence, that presents nonetheless real structural aspects of the physical world (Karakostas 2004b).

It is legitimate then to argue, in contrast to constructivist assertions, that the notion of 'mind-independent reality' is meaningful, useful and non-redundant for the following reasons: *Firstly,* the nonseparable character of mind-independent reality is scientifically established and not metaphysically guessed. Thus, quantum nonseparability, not being simply an artifact of human construction, represents a structural feature of the physical world that we positively know. *Secondly*, if one ignores the existence of mind-independent reality, he conceals the fact that human cognition aims at, and succeeds in, enlightening the actual character of the physical world. And, *lastly,* the recognition of the holistic character of mind-independent reality has led science to specify the conditions under which the maximum possible, empirically testable, information for the 'external' to the subject world could be *objectively* gathered and communicated. The last remark, properly understood, raises the requirement for a closer investigation of the issue of scientific objectivity in QM.

## 5.3. EMPIRICAL REALITY — A WEAK FORM OF SCIENTIFIC OBJECTIVITY

In the framework of QM, the 'epistemic' thesis of CSR (see § 3.3) ought to be revised. What contemporary physics can be expected to describe is not 'how mind-independent reality is', as CP may permit one to presume. The quantum mechanical formalism seems only to allow a detailed description of reality that is co-determined by the specification of a measurement context. The concept of 'active scientific realism', briefly presented at subsection 5.4 below, is just an implementation of such a contextualistic-realist interpretation of the theory. Within the domain of QM, the ideal conception of reality per se lacks a clear operational meaning. Instead, contextual quantum objects, acquiring a well-defined 'identity' within a concrete experimental context, constitute the 'scientific objects' of physical science. For, due to the genuinely nonseparable structure of QM and the subsequent context-dependent description of physical reality, a quantum object can produce no informational content that may be subjected to experimental testing without the object itself being transformed into a contextual object. Thus, whereas quantum nonseparability refers to an *inner-level* of reality, a mind-independent reality that is operationally inaccessible, the introduction of a context is related to the *outer-level* of reality, the contextual or empirical reality that results as an abstraction in the human perception (Karakostas 2003). In this sense, QM has displaced the verificationist *referent* of physics from 'independent reality' to 'empirical reality'. The latter is experimentally accessible, accurately describable, and scientifically objective.

It is important to note, however, that the meaning of the term 'objectivity' acquires, in general, different connotative status in the frameworks of classical and quantum mechanics. Within the domain of the latter, it is no longer possible to judge the objectivity of our knowledge through a comparison with the reality-itself, since the quantum description does not allow, in principle, complete knowledge of reality-itself. As argued in section 2, this state of affairs is intimately associated with the fact that, in contrast to CP, values of quantum mechanical physical quantities cannot be attributed to a microscopic object as intrinsic properties. The assigned values, being irreducibly contextual, cannot be regarded as *referring* to the object itself. This means that quantum mechanical statements concerning measurement or observation cannot be viewed as 'strongly objective', since they do not offer descriptions of how quantum entities 'actually are'. But should therefore be viewed as 'subjective' according to the constructivist claims? Certainly *not*, since, to start with, they are valid for everybody:

being built on the presupposition of subject-object separation, the observational statements of QM are perfectly harmonized with the perceptual limitations of all human beings. These statements do not describe the results of the subjective experience of human consciousness; they describe the 'manifestations' of nonseparable entangled physical phenomena, as these 'manifestations' emerge after the partition of the phenomena into 'observing system' and 'observed object' for the purposes of objectification or, at the laboratory level, of unambiguous communication. And this communication is ultimately obtained, not in virtue of the "relative compatibility of personally constructed concepts", as Glaserfeld argues, but in virtue of the uniform meaning assigned, by all human beings, to concepts fulfilling the condition of subject-object separation. We could thus say, in d' Espagnat's (1995) terms, that empirical reality obeys a weak form of objectivity, which secures *intersubjective agreement* with respect to QM observational statements.

The preceding analysis brings out the fact that *one and the same* quantum object does exhibit various possible contextual manifestations linked to the various possible incompatible observables pertaining to the object. Incompatible observables (such as the conjugated variables of position and momentum) acquire distinct contextual quantum states since, due to complementarity principle, they are physically well established in interaction to distinct, mutually exclusive, experimental arrangements. Contextuality is at the basis of quantum mechanical complementariry. Clearly, what the complementarity principle ultimately determines is the appropriate use of mutually exclusive objective descriptions (referring to mutually exclusive measurement contexts) permitting science to attain the maximal possible secure elucidation of the behaviour of a quantum system. This fact has led Bohr to conclude that "what we have learned in physics is how to eliminate subjective elements in the account of experience … we have fulfilled all requirements of an objective description of experience obtainable under specified experimental conditions" (Bohr 1955, pp. 568, 574). But if quantum mechanical contextuality secures scientific objectivity, is there any role left to the knowing subject itself?

## 5.4. ACTIVE SCIENTIFIC REALISM

The knowing subject — i.e., the experimenter/observer — is free to choose the aspect of nature he is going to probe, the question he selects to investigate, the experiment he is about to perform. The role of the experimenter's choice is ineliminable within the framework of QM. For, *firstly*, a choice on the part of the experimenter is not controlled or governed by the laws of quantum theory, *secondly*, the statistical algorithm of the theory can only be applied if a certain experimental context has first been selected, and, more importantly, *thirdly*, this element of freedom of choice in selecting the experimenter a context of discourse leads in QM to a gradually unfolding picture of reality that is not in general fixed by the *prior* part of the physical world alone. Thus QM calls for a kind of contextual realism that may be called 'active scientific realism' (Karakostas 2004a). *Active*, since it indicates the contribution of rational thought to experience; it acknowledges the *participatory* role of the knowing subject in providing the context of discourse; as has already been noted, the identification of a specific pattern as an 'object' depends on the process of knowledge, on the minimization or suppression of the entangled correlations between the 'object' concerned and its external environment, and this may be done in ways depending on the chosen context of investigation. And *realism*, since given a context, concrete objects (structures of reality) have well-defined properties independently of our knowledge of them.

Even if our knowledge will never be knowledge of reality-itself, independent of the way it is contextualized, there is no reason, from the viewpoint of natural philosophy, to reject as meaningless the idea of such a mind-independent reality. On the contrary, the conception of active scientific realism, by acknowledging the subject-object inherent engagement that is built into QM, attempts to incorporate the human factor into our proper understanding of reality. In this respect, the actual relation of the knowing subject to a world should neither be viewed as a relation of absolute dichotomy, nor of independence or of externality, but of active participancy. As far as knowledge acquisition is concerned, the latter relation must be understood as involving a two-way mutual epistemological process: thus in coming to know the physical world, we expect to know also how our perspective affects or contributes to our conceptualization about the world, since in contemporary physics such an effect is unavoidable; and conversely, by knowing how we contribute to the knowledge claims about the world, we more securely identify the informational content the world itself contributes.

The foregoing analysis discloses the fact that the ontological scheme of CSR is incompatible with QM worldview. The source of the incompatibility is located in the generalised phenomenon of quantum mechanical nonseparability, which indisputably contradicts the 'independence dimension' of CSR. In recapitulation, it is worth to notice the following points: *Firstly*, QM questions the 'metaphysical' thesis of CSR, since it reveals the active participation of the knowing subject in providing the context of investigation; the outer reality is no longer perceived as something given a priori, a 'ready-made' truth, passively discovered, but as something affected by the subject's action. *Secondly,* QM also questions the 'semantic' thesis of CSR, since its validity presupposes a specific assumption about the structure of the world, namely that the world consists or is built out of well-defined and independently existing objects. As already has been extensively argued, however, from the viewpoint of modern quantum theory, any a priori identification of 'physical objects' with 'physical reality' is inadmissible, because — whatever the precise meaning of 'physical objects' may be — we have to expect that such systems, according to the theory, are entangled by nonseparable correlations, so that they lack intrinsic individuality, autonomous well-defined existence. And, *lastly*, the 'epistemic' thesis of CSR is undeniably shaken, since in the framework of QM, scientific theories cannot be viewed as accurate representations of the world as 'it really is'; due to quantum nonseparability, physical reality considered as a whole is not scientifically completely knowable; any detailed description of it necessarily results in irretrievable loss of information by dissecting the otherwise undissectable. Hence, our knowledge claims to reality can only be partial, not total or complete, extending up to the structural features of reality that are approachable through the penetrating power of our theoretical apparatus.

The nonseparable structure of QM, the experimentally well-confirmed entangled features arising out of it, and the subsequent context-dependent description of physical reality conflict with the rather comfortable view that the world's contents enjoy an absolute identity of existence. Yet, scientific realism and quantum nonseparability are *not* incompatible. The relationship between them points, however, to the abandonment of the classical conception of physical reality and its traditional metaphysical assumptions. Thus, by defending a new form of realism in contemporary physics, we attempted indeed to show that at the constructivist spectrum certain of its claims appear at least unconvincing. To this end, we recall, for instance, that it seems hardly consistent to dispense completely with any notion of reality that is logically *prior* to experience and knowledge. But how could all these contribute to QM instruction? Let us roughly sketch our educational strategy.

## 6. The Contribution of History and Philosophy of Science on Quantum Mechanics Instruction: An Educational Proposal

The project, in which this study is embedded, attempts to formulate a theoretical instructional intervention that may gradually lead learners to a proper discrimination of the conceptual assumptions that underlie the frameworks of classical and quantum physics. To this end, it takes critically into account and often embraces conceptual change strategies suggested by educational constructivism. But this fact in no way prevented us to adopt a series of objections raised against constructivist epistemology (e.g., an extended critical examination is included in Matthews 1998). We observed, for instance, that the constructivist idea concerning the sensorial or social origin of scientific knowledge often results to the *downplaying* or *neglect* of scientific content. In relation to educational matters, this is tightly associated of course with the constructivist perception that the main aim of science education is to enable students to function effectively in their personal or social world.

On the contrary, we take the view that the scope of science education is to lead students to 'reconstruct' their initial knowledge through a process that *has* to be performed under the *close guidance* of the teacher who attempts to communicate what has recently been called the "Nature of Science" (Mc Comas 2000). The content of this term, *firstly,* includes the contemporarily accepted body of scientific knowledge, *secondly,* reflects the basic aspects of scientific thinking and reasoning, and, *lastly,* mirrors the historical development of scientific knowledge. It is evident that, from this point of view, science learning is evaluated by reference to the currently accepted scientific content, the norms, the values and the methodology of science, and *not* by mere reference to its personal or social 'utility'. Furthermore, and in a reverse way, this point of view permits also science education to draw valuable instructional tools from the domains of history and philosophy of science for overcoming learners' cognitive difficulties (Hadzidaki 2003).

Following this line, we have formulated an epistemological structure, labeled "Levels of Reality", which constitutes the base of the proposed educational strategy (Hadzidaki, Kalkanis & Stavrou 2000). This structure has been shaped under our premise (analysed in the introduction) that learners' misconceptions concerning quantum phenomena should be viewed, in virtue of their empirically established *persistence,* as 'epistemological obstacles' (in the sense of Bachelard 1938) to the acquisition of QM knowledge. This premise, associating by its very nature the intended conceptual reconstruction of learners with the development of scientific knowledge, becomes the link between the *cognitive* and *epistemological* foundation of the instructional process. More specifically, the 'Levels of Reality' structure, aiming initially at securing an effective discrimination between the content of successive scientific theories, suggests *a context approach* to physics' teaching. Physical phenomena are broadly classified into distinct 'Levels' (e.g., the 'Classical Physics Level' and the 'Quantum Physics Level'), which represent independent conceptual systems of the corresponding theories. But intending also to make visible the non-fragmented or even unified image of the natural world accepted by modern physics, the 'Levels of Reality' structure links the distinct 'Levels' through the appropriate reconstruction of material drawn from the domains of history and philosophy of science. The instructional 'simulation' of the processes that have led science itself to theory change offers, in our opinion, a plausible account of the meaning and structure of the present body of scientific knowledge, an account that also induces understanding (Hadzidaki & Karakostas 2001). In this respect, and in connection with our preceding considerations, students are expected to realize that the claims of CSR — representing to a large extent their own intuitive metaphysical beliefs — shape essentially the ontological background of the 'Classical Physics Level'. Thus, the ontological scheme of CSR *is juxtaposed* with a perception of physical reality, which, by moving in the direction of 'active scientific realism', reveals the inner meaning of the concepts included in the 'Quantum Physics Level'. Such a process, responding to the requirements of cognitive psychology (see introduction), is expected to support learners' conceptual change towards an appropriate QM worldview. It remains of course to the empirical implementation of this proposal to reveal its real perspectives as well as its possible limitations.